\documentclass[aps,pra,showpacs,twoside,twocolumn,10pt,longbibliography,floatfix]{revtex4-2}
\usepackage[colorlinks=true, citecolor=red, urlcolor=blue ]{hyperref}
\usepackage{times,epsfig,amssymb,amsfonts,amsmath,subfigure,mathtools,amsthm,braket,soul,enumitem,color}
\usepackage[normalem]{ulem}

\newcommand{\stkout}[1]{\ifmmode\text{\sout{\ensuremath{#1}}}\else\sout{#1}\fi}
\usepackage{cases}
\usepackage{amsmath}
\usepackage{dsfont}
 \usepackage{relsize}
\usepackage{bm}
\usepackage[dvipsnames]{xcolor}

\makeatletter
\DeclareRobustCommand*{\bfseries}{%
  \not@math@alphabet\bfseries\mathbf
  \fontseries\bfdefault\selectfont
  \boldmath
}
\makeatother


	\definecolor{blue-violet}{rgb}{0.54, 0.17, 0.89}

\begin{document}
\raggedbottom




\title{Temperature- and interaction-tweaked efficiency boost of  finite-time robust quantum Otto engines}

\author{Debarupa Saha, Ahana Ghoshal, and Ujjwal Sen}
\affiliation{ Harish-Chandra Research Institute, A CI of Homi Bhabha National Institute, Chhatnag Road, Jhunsi, Prayagraj 211 019, India}

\begin{abstract}
We demonstrate that under specific conditions, a finite-time quantum Otto engine, employing a spin-$1/2$ particle as the working substance, despite undergoing incomplete Otto cycles, can achieve higher efficiency than an ideal quantum Otto engine. A finite-time quantum Otto engine refers to an Otto engine where the two isochoric strokes are prematurely terminated before reaching thermal equilibrium with their respective hot and cold baths. This contrasts with the ideal quantum Otto engines, which operate with perfect isochoric strokes, establishing thermal equilibrium between the system and the corresponding bath.  We observe that the enhancement of efficiency of a finite-time quantum Otto engine over the ideal one can be realized by
adjusting the initial temperature of the working substance
within the temperature range of the hot and cold baths. We also find that incorporating an auxiliary qubit,
and activating specific interactions between the single-qubit working substance and the auxiliary one, can enhance the efficiency of a finite-time as well as an ideal quantum Otto engine. Furthermore, we analyze the impact of glassy disorder within the system-bath coupling during the two isochoric strokes on the efficiency of a finite-time quantum Otto engine. Our findings reveal that as the strength of disorder increases, the efficiency of a finite-time quantum Otto engine tends to decrease, albeit with a relatively modest reduction even at high disorder strengths. However, the advantage in efficiency of the finite-time quantum Otto engine over the ideal one, obtained by tuning the initial state temperature, and the efficiency enhancement obtained by incorporating an auxiliary qubit over the without- auxiliary scenario, persists even in the presence of substantial disorder. Additionally, we show that while this disorder does not have an impact on the ideal efficiency, it does influence the duration of isochoric strokes needed for a quantum Otto engine to reach its ideal efficiency. This stroke duration remains nearly constant up to a specific value of disorder strength. Beyond that point, as disorder strength increases, the required duration of time 
increases rapidly.

\end{abstract}
\maketitle
\section {Introduction}
The miniaturization of thermal devices, in particular for its efficient use in quantum technologies, is one of the reasons that has led to the development of quantum heat engines, designed to convert heat energy into mechanical energy within the quantum realm~\cite{R_Alicki_1979,kosloff1984quantum,Allahverdyan2008,quan2009quantum,Scully2011,Fialko2012,Kosloff2014,Abah2014,Niedenzu_2016,Uzdin2016,Friedenberger_2017,Niedenzu2018}. The operation of a quantum heat engine is governed by  quantum versions of the familiar thermodynamic cycles, like Carnot cycle~\cite{Carl_M_Bender_2000,Bender2022,Alicki2013,Abiuso2020}, Otto cycle~\cite{henrich2007quantum,quan2007quantum,abah2012single,Uzdin_2014,Leggio2016,Kosloff2017,mehta2017quantum,Barrios2017,Solfanelli2020,Grosso2022}, etc. In classical thermodynamics, ideal gases typically serve as the working substance in heat engines. However, in the quantum regime, the quantumness of a heat engine is, in general, incorporated in the working substance, often having discrete energy levels, and in its dynamics~\cite{Quan2005,Latifah2011,Latifah2013}. Over the past few decades, the proposals for fabrication of quantum heat engines has seen various quantum-mechanical models employed, including spin-$1/2$ particles~\cite{He2002,feldmann2004characteristics,henrich2007quantum,Kieu2004,Thomas2017,Watanabe2017}, three-level quantum systems~\cite{Scovil1959,Geva1996,Li_2007}, potential wells~\cite{Carl_M_Bender_2000,Quan2005,Wu2010,Latifah2011,Francisco2012,Latifah2013,Purwanto2016,Michel2016}, harmonic oscillators~\cite{rf1, Kosloff2017,rezek2006irreversible,he2009performance,Insinga2016}, ions in harmonic traps~\cite{Kosloff2017}, multiferroic chains~\cite{Azimi2014}, etc. Some of these models have been successfully realized in experimental setups involving trapped ions~\cite{maslennikov2019quantum,PhysRevLett.123.080602}, NMR~\cite{PhysRevLett.123.240601}, NV centers~\cite{PhysRevLett.122.110601}, superconducting qubits~\cite{PRXQuantum.2.030353}, ultra cold atoms~\cite{bouton2021quantum}, and others. The operation and efficiency of quantum heat engines can be influenced by various quantum-mechanical phenomena, such as quantum coherence~\cite{Scully2011,Rahav2012,Uzdin2015,Korzekwa_2016,Brandner2017,Camati2019,Dorfman2018}, quantum entanglement~\cite{Zhang2007,Funo2013} and other types of quantum correlations~\cite{altintas2014quantum,Barrios2017RoleOQ,xiao2022photo} and interactions~\cite{asadian2022quantum,piccitto2022ising} 

Quantum Otto engines, whose operation is based on the principles of quantum Otto cycles, is the central focus of this paper. The working substance of such an engine undergoes four distinct strokes, comprising two isochoric and two adiabatic processes. In each of the isochoric strokes, the working substance, which is a quantum system, remains in contact with a thermal reservoir, and the time dynamics of the system is governed by the corresponding open quantum evolution~\cite{davies1974markovian,Breuer2002}. In contrast, in the adiabatic strokes, the working substance undergoes slow modifications to its Hamiltonian, typically without any heat exchange with the surrounding environment. 
The efficiency of any heat engine is classically bounded by the Carnot efficiency, as derived by S. Carnot~\cite{Car}. This efficiency bound applies not only to classical heat engines but also to quantum ones, including  quantum Otto engines, providing a fundamental limit on their performance~\cite{jahnke2008nature,ref2}. 
This bound is typically achievable by an ideal reversible engine, a scenario that often necessitates an infinite time scale of the isochoric strokes and results in vanishing power ~\cite{VP1,VP}. However, the requirement of 
an infinite-time evolution unreachable in practical scenarios. Practical heat engines operate in finite timescales, delivering finite-power outputs and thereby leading to irreversible processes. Consequently, due to practical considerations and in the pursuit of enhanced power output, the study of heat engines has increasingly focused on the finite-time regime and on performance far from equilibrium~\cite{mehta2009performance,abah2019shortcut,wang2019finite}. In this realm of irreversible finite-time heat engines, the maximum power efficiency is classically constrained by the Curzon-Ahlborn efficiency ~\cite{Bjarne1977,Broeck2005}. While some working substances within finite-time quantum Otto engines adhere to this classical limit~\cite{geva1992classical,Esposito2009}, it is noteworthy that, under specific assumptions, quantum heat engines have demonstrated the capability to surpass this classical constraint~\cite{Deffner2018}.
Various finite-time quantum Otto engines have been examined in the literature, employing diverse configurations such as two-level systems~\cite{Geva1992,feldmann2000performance}, harmonic oscillators~\cite{rf1,geva1992classical,Kosloff2017}, coupled harmonic oscillators~\cite{Wang2015}, squeezed thermal bath~\cite{Wang2015}, etc. See also~\cite{geva1994three,wu2014efficiency, chen2019boosting,das2020quantum,lee2020finite,saryal2021bounds,barrios2021light} in this regard. For more works on quantum Otto engines see \cite{wang2009thermal,thomas2011coupled,he2012thermal,zagoskin2012squeezing,altintas2015quantum,chen2019boosting,Chand_2017,kloc2019collective,camati2019coherence}.

Impurities and fluctuations, stemming from the erroneous implementation of quantum devices or the influence of environmental factors, or even a combination of both, can give rise to a disordered quantum system. Such a disordered system often exhibits behaviors that deviate from the expectations of a perfectly ordered one. In general, disorder tends to disrupt the naturally advantageous behaviors of quantum systems, leading to reduced efficiency in quantum devices. However, in some cases, disorder has been shown to provide benefits compared to the ordered situations~\cite{aharony1978spin,santos2004entanglement,niederberger2008disorder,prabhu2011disorder,martin2014quenched, diep2013frustrated}. In the context of a quantum Otto engine, disorder can manifest, e.g. when tuning the parameters of the working substances and the baths within the composite setup. Additionally, the working substances can be susceptible to environmental influences, leading to oscillations or drifts in the system. When disorder is present, three possible scenarios emerge: (i) the efficiency of the engines may decrease, (ii) the quantum Otto engines could display robustness against the presence of disorder, or (iii) disorder might even offer advantages over the ideal setup. In the latter two cases, disorder can prove beneficial in the practical implementation of a quantum Otto engine. Therefore, investigating the effects of disorder on the quantum Otto engines holds significant importance. See e.g.~\cite{alecce2015quantum} in this regard.

In this paper, our main focus is to study the operation characteristics of a finite-time quantum Otto engine in ordered and disordered scenarios. Such an engine involves premature termination, possibly due to practical limitations, of two isochoric strokes before reaching thermal equilibrium with the hot and cold baths, which is in contrast with an ideal quantum Otto engine that execute perfect isochoric strokes, ensuring thermal equilibriums with the corresponding baths. In this work, we specifically consider a quantum Otto engine employing a quantum spin-$1/2$ particle as its working substance.  We observe that, even while encountering incomplete Otto cycles, these single-qubit finite-time quantum Otto engines, in specific scenarios, can achieve higher efficiency than their ideal counterparts. We find that the efficiency enhancement of a finite-time quantum Otto engine can be accomplished 
by fine-tuning  the initial temperature of the working substance within the temperature range defined by the hot and cold baths. Also, by introducing an auxiliary qubit and enabling specific interactions of it with the system qubit, which is the spin-$1/2$ particle employed as the working substance, the efficiency of finite-time as well as an ideal quantum Otto engine can be improved. Furthermore, we have explored the influence of disorder in the system-bath coupling, specifically during the two isochoric strokes, on the efficiency of a finite-time quantum Otto engine. We show that with increasing disorder strength, the efficiency of the finite-time quantum Otto engine experiences a reduction, albeit remaining relatively moderate, even at high strengths of disorder. Significantly, the efficiency advantage retained by the finite-time quantum Otto engine over its ideal counterpart, obtained by tuning the initial temperature of the working substance,
remains intact even in situations where substantial disorder is present. And, the efficiency enhancement observed in finite-time quantum Otto engines, resulting from the incorporation of an auxiliary qubit over the without auxiliary setup, remains effective under similar conditions of substantial disorder. Our study also demonstrates that the presence of this type of disorder does not impact the ``ideal efficiency", the efficiency of an ideal quantum Otto engine. However, it does affect the duration of the isochoric strokes in a quantum Otto engine needed to reach this ideal efficiency. This time duration maintains a nearly constant value until a specific level of disorder strength is reached. Beyond this threshold, an increase in disorder strength is accompanied by a rapid increase in the required stroke duration.


The remainder of the paper is organized as follows. In Sec.~\ref{sec:2}, we provide a description of the four strokes involved in a finite-time quantum Otto engine that utilizes a spin-$1/2$ particle as the working substance. Section~\ref{sec:3} delves into the efficiency of a finite-time quantum Otto engine. We explore how adjustments to the temperature of the initial state and an incorporation of a transverse field impact the engine's efficiency within this section. In Sec.~\ref{sec:4}, we introduce an auxiliary qubit alongside the working substance and analyze its influence on the efficiency of a finite-time quantum Otto engine.
Section~\ref{sec:5} contains the effects stemming from glassy disorder within the system-bath coupling, particularly during the isochoric strokes, on the efficiency of a finite-time quantum Otto engine. In Sec.~\ref{sec:6}, we present our concluding remarks and enumerate the key findings of this study.
\section{Finite-time quantum Otto engine}
\label{sec:2}
\begin{figure*}
    \centering
    \includegraphics[width=8.5cm]{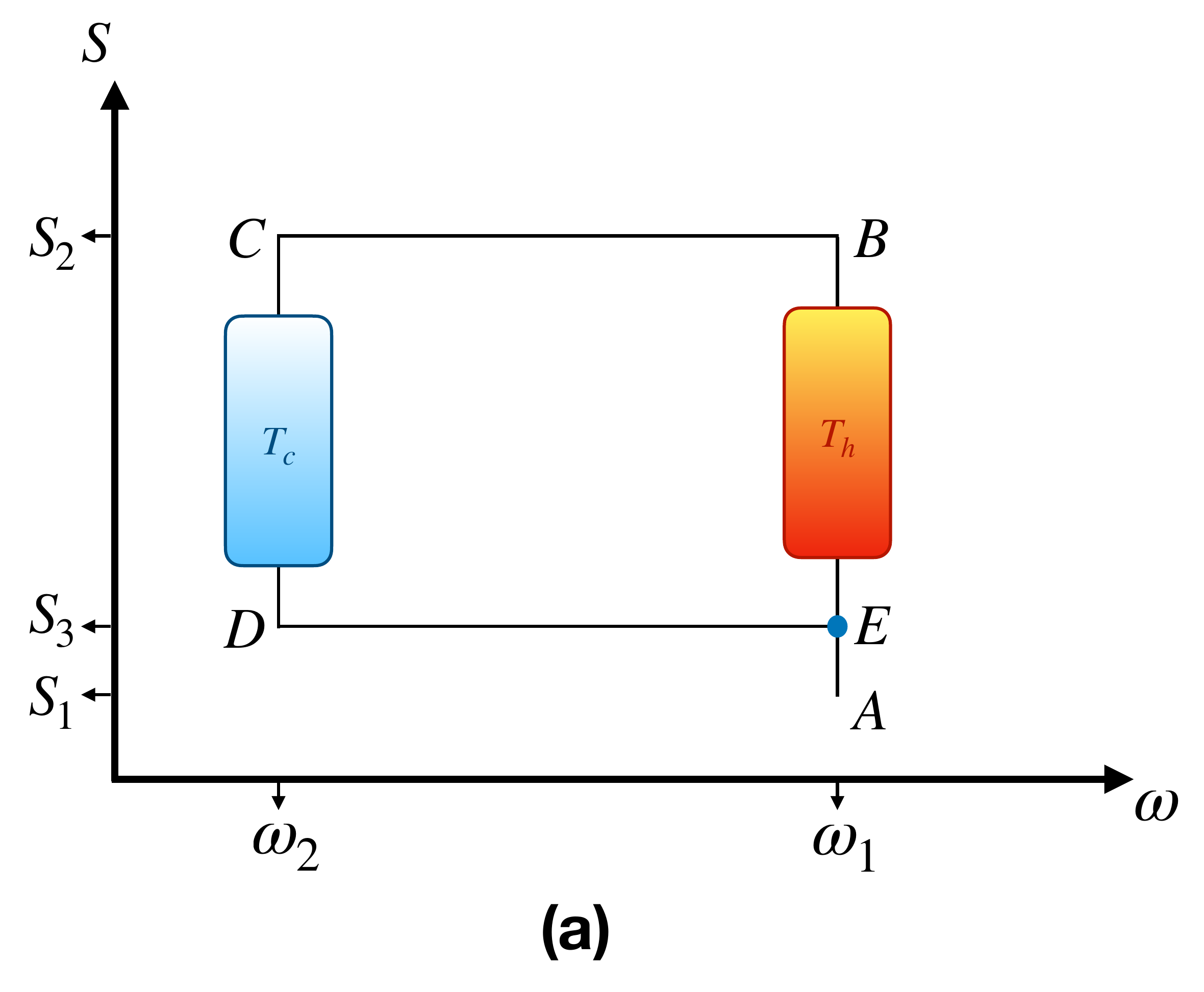}
    \includegraphics[width=8.5cm]{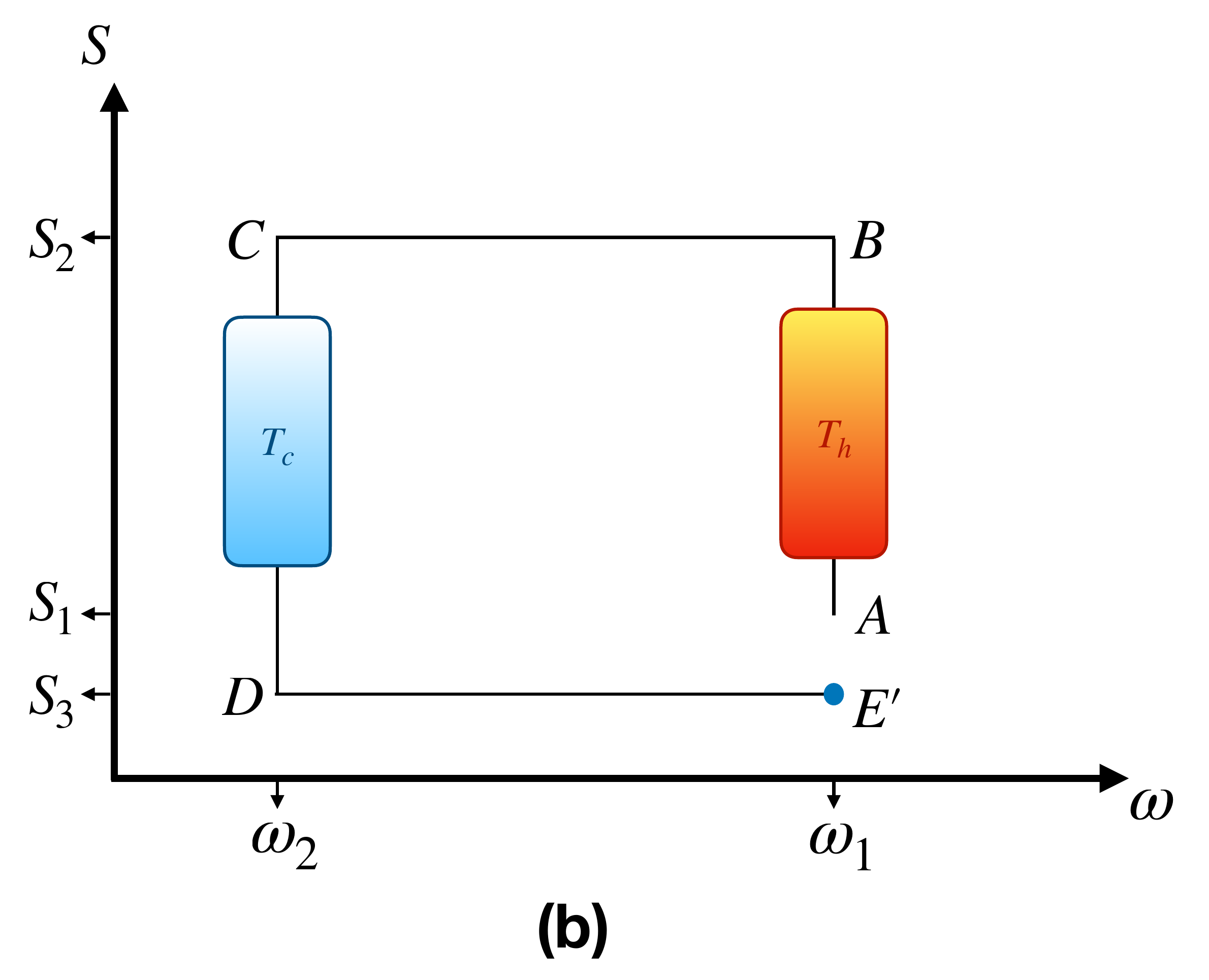}
    \caption{Schematic diagram of a finite-time four-stroke quantum Otto engine. Here we depict the four consecutive strokes of a finite-time quantum Otto engine in the magnetization ($S$) vs. frequency ($\omega$) plane. The paths $A\rightarrow B$ and $C\rightarrow D$ represent two isochoric processes in which the system is in contact with hot and cold baths at temperatures $T_h$ and $T_c$, respectively. The other two paths, $B\rightarrow C$ and $D\rightarrow E$ or $E^{\prime}$, represent two adiabatic strokes in which the evolution of the system is governed by unitary transformations. As an effect of the finite time isochoric strokes, in which the evolution of the system is truncated before reaching the thermal equilibrium state, the Otto cycle remains incomplete. Hence, two possible incomplete Otto cycles can occur and are presented in panels (a) and (b). 
    In panel (a), the final magnetization of the system, denoted as $S_{3}$, exceeds the initial magnetization $S_{1}$. Conversely, in panel (b), we depict a situation where $S_{3}<S_{1}$.  
    In the path $B\rightarrow C$ the frequency associated with the system changes from $\omega_1$ to $\omega_2$ and exactly the reverse situation occurs during the path $D \rightarrow E$ for panel (a) and $D\rightarrow E^{\prime}$ for panel (b).}
    \label{fig:fig1}
\end{figure*}
As mentioned previously, akin to an ideal quantum Otto engine, a finite-time quantum Otto engine, incorporating a quantum mechanical working substance, also undergoes a sequence of four consecutive strokes. During the first stroke, it absorbs heat from a hot thermal bath, symbolizing an isochoric heating process, whereas in the third stroke, it releases heat into a cold bath, representing an isochoric cooling process. The remaining two strokes involve adiabatic expansion and compression processes, both governed by unitary evolution of the working substance. In this paper, we denote the temperature of the hot bath as $T_h$ and the temperature of the cold bath as $T_c$, where $T_h > T_c$, and the thermal baths involved in the two isochoric strokes are taken as bosonic baths, each consisting of an infinite number of harmonic oscillators. While various multi-level quantum systems have been explored as working substances in the existing literature, this paper exclusively focuses on a spin-$1/2$ system as the chosen working substance. 
A schematic diagram illustrating the four-stroke finite-time quantum Otto engine is provided in Fig.~\ref{fig:fig1}, represented in the magnetization ($S$) versus frequency ($\omega$) plane. Here, the magnetization $S$ is defined as $S = \langle S_z \rangle = \langle \frac{1}{2}\sigma_{z} \rangle = \text{Tr}[\rho S_{z}]$, where $\sigma_z$ represents the Pauli-$z$ matrix, and $\rho$ denotes the density matrix of the system. The frequency, denoted as $\omega$, corresponds to the frequency associated with the Hamiltonian of the system, given by $H_{\mathcal{S}}=\frac{1}{2}\hbar\omega\sigma_z$.

The Otto cycle begins at point $A$ (as shown in Fig.~\ref{fig:fig1}). We initiate the cycle with a spin-$1/2$ particle, prepared in the state $\rho_A$, which is in thermal equilibrium with the cold bath at temperature $T_c$ corresponding to the Hamiltonian $H_{\mathcal{S}_2} = \frac{1}{2}\hbar\omega_2\sigma_{z}$, i.e., $\rho_A = \frac{e^{-\beta_c H_{\mathcal{S}_2}}}{Z_2}$, where $Z_2 = \text{Tr}(e^{-\beta_c H_{\mathcal{S}_2}})$, and $\beta_c = \frac{1}{k_BT_c}$, with $k_B$ being the Boltzmann constant. 
Here $\omega_2$ represents the strength of the magnetic field employed during the state preparation. The magnetization of the system at point $A$ is denoted as $S_1$, calculated as $S_1 = \text{Tr}[\rho_A S_z]$. The subsequent four strokes of the cycle are discussed below.
\subsection{The first stroke: isochoric heating ($\mathbf{A\rightarrow B}$)} 
The path from point A to point B in both panels of Fig.~\ref{fig:fig1} represents the quantum isochoric heating process and is considered as the first stroke of a conventional heat engine. During this stroke, the system is immersed in a thermal reservoir (the hot bath), which is kept in a state of thermal equilibrium at a temperature denoted as $T_h$. 
Throughout this process, the system undergoes an open quantum evolution in the presence of the hot bath. Importantly, during this evolution, the Hamiltonian of the system remains constant and is defined as $H_{\mathcal{S}_1}=\frac{1}{2}\hbar\omega_1\sigma_z$.
This entire process is quantum isochoric in the sense, that the strength of the applied field $\omega_1$ is held constant throughout the evolution. To describe a general isochoric process, let us consider the working substance, i.e., the spin-$1/2$ particle, is attached to a thermal reservoir at temperature $T_i$. 
The total Hamiltonian of the composite system-bath setup can be now expressed as 
\begin{equation}
\label{hamil}
    H=H_{\mathcal{S}_1}+H_{\mathcal{B}_i}+H_{\mathcal{SB}_i},
\end{equation}
where the free Hamiltonian of the reservoir is defined as
 \begin{equation}
H_{\mathcal{B}_i}=\sum_{k}\hbar\omega^{\prime}_{k} a_{k}^{i\dagger} a_{k}^i,
 \end{equation}
and the system-bath interaction is given by 
\begin{equation}
\label{int}
H_{\mathcal{SB}_i}=\sigma_{x}\sum_{k}\hbar g_{k}^i(a_{k}^{i\dagger} + a_{k}^{i}).
\end{equation}
Here $\omega_k^{\prime}$ represents the frequency of the $k^{\text{th}}$ mode, and $a_k^i$ ($a_k^{i\dagger}$) denote the bosonic annihilation (creation) operators corresponding to the $k^{\text{th}}$ mode of the harmonic oscillator bath. The term $\sigma_{x}$ represents the Pauli-$x$ matrix operating on the Hilbert space of the system. 
Additionally, $g_{k}^i$, having the unit of frequency, determines the coupling strength of the interaction between the system and the $k^{\text{th}}$ mode of the reservoir.
Here, it is assumed that the coupling between the system and the reservoir is sufficiently weak, leading to a Markovian dynamics of the system, which obeys the 
the Born-Markov and secular approximations
~\cite{Breuer2002,vadimov2021validity}. Consequently, the reduced system dynamics from point $A$ to point $B$, as depicted in Fig.~\ref{fig:fig1}, is governed by the Gorini-Kossakowski-Sudarshan-Lindblad (GKSL) master equation, given by,
\begin{equation}
\label{GKSL}
    \frac{d\rho(t)}{d(\tilde{\omega} t)}=-\frac{\iota}{\hbar\tilde{\omega}}[H_{\mathcal{S}_1},\rho(t)]+\frac{1}{\tilde{\omega}}\mathcal{L}_i[\rho(t)], 
\end{equation}
with $\tilde{\omega}$ being a constant with the unit of frequency. Therefore, the term $\tilde{\omega}t = t^{\prime}$ serves as the dimensionless time parameter. Moreover, the expression $\mathcal{L}_i[\rho(t)]$ denotes the dissipative term originating from the environmental influence, and it is described as
 \begin{align}
 &\mathcal{L}_i[\rho(t)]=\nonumber\\
 &\phantom{a}\sum_{\mathcal{E}}\gamma_i(\mathcal{E})[A(\mathcal{E})\rho(t) A^\dagger(\mathcal{E})
- \frac{1}{2}\{A^\dagger(\mathcal{E}) A(\mathcal{E}),\rho(t)\}]
\end{align}
with the Lindblad operators,
\begin{equation}
    A(\mathcal{E})=\sum_{\mathcal{E}=\epsilon_r-\epsilon_l}\ket{l}\bra{l}\sigma_{x}\ket{r}\bra{r},
\end{equation}
where $\ket{l}$ and $\ket{r}$ are the eigenvectors of the system Hamiltonian $H_{\mathcal{S}_1}$ corresponding to the energy eigenvalues $\hbar \epsilon_l$ and $\hbar \epsilon_r$ respectively. $\gamma_i(\mathcal{E})$ are the transition rates corresponding to the transition energies $\hbar\mathcal{E}$, defined as
\begin{align}
\label{gamma}
\gamma_i(\mathcal{E})=
    \begin{cases}
        \mathcal {J}_i(\mathcal{E})(1+\overline{n}_i(\mathcal{E})), & \mathcal{E} > 0,\\
        \mathcal {J}_i(|\mathcal{E}|)(\overline{n}_i(|\mathcal{E}|), & \mathcal{E} < 0.
    \end{cases}
\end{align}
Here, $\mathcal{J}_i(\mathcal{E})$ represents the ohmic spectral density of the bosonic bath. We define it as $\mathcal{J}_i(\mathcal{E}) = \frac{g_{\mathcal{E}}^{i^2}}{\tilde{\omega}} = \lambda_i\mathcal{E}$, where $\lambda_i$ is a dimensionless constant.
$\overline{n}_i(\mathcal{E})$ corresponds to the Bose-Einstein distribution function, and it is given by $\overline{n}_i(\mathcal{E}) = \frac{1}{\exp\left(\frac{\hbar\mathcal{E}}{k_{B}T_{i}}\right)-1}$. As, in this stroke, the system is coupled to the hot reservoir with temperature $T_h$, the $i$ will be replaced by $h$ in all the above expressions.
If we allow the system to evolve for a sufficiently long time under the influence of the hot bath obeying the Born-Markov and secular approximations~\cite{Breuer2002,vadimov2021validity}, it will eventually reach the canonical equilibrium state at temperature $T_h$. However, in the context of this paper, our primary focus lies on exploring the finite-time behavior of quantum Otto engines. Therefore, we need to concentrate our attention on the transient regime of the evolution, which occurs before the system attains its canonical equilibrium state.    
Let us consider that we truncate the open quantum evolution of the system before it reaches equilibrium, precisely at the dimensionless time $t^{\prime}=\tilde{t}$, at point $B$ (see Fig.~\ref{fig:fig1}). Suppose, at this point, the final state of the system is $\rho_B$. Throughout this Markovian evolution from point $A$ to point $B$, the occupation probabilities of various energy levels change with time. So, this results in a variation in the magnetization as time progresses. Let us assume that at point $B$, the magnetization of the system is quantified as $S_2=\text{Tr}[\rho_B S_z]$.
However, it is crucial to note that during this stroke, the magnetic field remains constant. Therefore, the work done from point $A$ to point $B$ amounts to zero. Hence, the total heat, denoted as $Q_1$, extracted from the heat bath by the system is equal to the change in the internal energy of the system. In other words, we have
\begin{equation}
    Q_1= \text{Tr}[H_{\mathcal{S}_1}(\rho_{B}-\rho_{A})].
\end{equation}
  \subsection{The second stroke: adiabatic compression ($B\rightarrow C$)}
  \label{sec:2.B}
The second stroke of a quantum Otto engine is characterized by a quantum adiabatic compression process, during which the system is isolated from its surrounding environments, and the strength of the magnetic field is gradually reduced from $\omega_{1}$ to $\omega_{2}$ along the path from point $B$ to point $C$, as depicted in both the panels of Fig.~\ref{fig:fig1}. Typically, this process is governed by a unitary evolution, and in this paper, we adopt the driving Hamiltonian for this unitary evolution as 
\begin{equation}
\label{driving1}
    H_{\mathcal{S}}(t)=\frac{1}{2}\hbar\omega(t)\sigma_z,
\end{equation}
where $\omega(t)=\omega_1(1-\frac{t}{\tau})+\omega_2(\frac{t}{\tau})$, and $\tau$ is an arbitrary but fixed time. At $t=0$, $H_{\mathcal{S}}(0)=H_{\mathcal{S}_1}$ and at $t=\tau$, $H_{\mathcal{S}}(\tau)=H_{\mathcal{S}_2}$. Consequently, the evolution of the system from point $B$ to point $C$ is dictated by the unitary evolution expressed as
\begin{equation}
    \rho_{C}=U\rho_{B}U^\dagger,
\end{equation}
where $U$ is given by,
\begin{equation}
   U= e^{-i\int_{0}^{\tau} H_{S}(t)\,dt}.
\end{equation}
The quantum adiabatic process is a thermodynamic phenomenon in which the population of the energy levels within the system remains unchanged as time progresses. This condition can only be satisfied if the process unfolds at an infinitesimally slow pace, necessitating $\tau$ to be infinitely large, but achieving such extreme slowness in practical situations can be challenging. Therefore, various methods have been introduced in prior literature to facilitate and attain adiabaticity~\cite{messiah2014quantum,hwang2015quantum,srivastava2020scaling}.
For this work, given that the initial state $\rho_A$ is diagonal in the energy eigenbasis of $H_{\mathcal{S}_2}$ and $[H_{\mathcal{S}_1},H_{\mathcal{S}_2}]=0$, after the Markovian evolution from $A\rightarrow B$, the state of the system at point $B$ will also be diagonal in the eigenbasis of $H_{\mathcal{S}_1}$. Thus, we can express the state at point $B$ as $\rho_B=\sum_{m} P^{\prime}_{B_{m}}\ket{m(B)}\bra{m(B)}$, where $\ket{m(B)}$ represents the eigenvectors of the Hamiltonian $H_{\mathcal{S}_1}$.
Now, in case of a perfect adiabatic process, during the path $B \rightarrow C$ the populations of the energy levels will not change. Therefore, at point $C$, the state of the system will be $\rho_C=\sum_{m} P^{\prime}_{B_{m}}\ket{m(C)}\bra{m(C)}$, where $\ket{m(C)}$ are the eigenvectors of the Hamiltonian $H_{\mathcal{S}_2}$. 
As the populations of the energy levels do not change during this adiabatic compression, the magnetization also remains constant at $S_2$ during the path from $B$ to $C$.
Since the system is isolated from its surroundings during this stroke, no heat is exchanged. However, there is a change in the internal energy of the system due to the alteration in the magnetic field, which is equivalent to the amount of work done. Thus, the work done during the path from $B$ to $C$ is given by, 
\begin{equation}
    W_{1}=\text{Tr}[H_{\mathcal{S}_1}\rho_{B}-H_{\mathcal{S}_2}\rho_{C}].
\end{equation}

\subsection{The third stroke: isochoric cooling ($C \rightarrow D$)}
Following the adiabatic compression, as the system reaches point $C$, it is once again connected to an external reservoir, but this time, the reservoir is colder, with a temperature denoted as $T_c$, as depicted in the two panels of Fig.~\ref{fig:fig1}. 
During this stroke, the magnetic field strength is kept constant at $\omega_2$. Therefore, the 
evolution of the system in presence of this cold reservoir will be governed by the same GKSL master equation as given in Eq.~(\ref{GKSL}), with the first term involving $H_{\mathcal{S}_1}$ should be replaced by $H_{\mathcal{S}_2}$ and $i$ in all the expressions will be replaced by $c$.
Additionally, the quantities $\epsilon_l$ and $\epsilon_r$, as well as the vectors $\ket{l}$ and $\ket{r}$ used for defining transition energies $\mathcal{E}$ and constructing the jump operators $A(\mathcal{E})$, should now correspond to the eigenfrequencies and eigenvectors of $H_{\mathcal{S}_2}$ respectively, rather than those of $H_{\mathcal{S}_1}$. 
In this stroke, as the system is connected to a colder bath, it releases heat to the bath and undergoes a transformation from $\rho_C$ to $\rho_D$. Consequently, the magnetization changes from $S_{2}$ to $S_{3}=\text{Tr}[\rho_{D}S_{z}]$. The amount of heat released in this stroke thus can be expressed as
 \begin{equation}
    Q_{2}=\text{Tr}[H_{\mathcal{S}_2}(\rho_{D}-\rho_{C})].
\end{equation}
Throughout the entire discussion of this paper, for the two isochoric strokes, we set $\lambda_h=\lambda_c=0.1$. This choice ensures that the coupling between the system and the reservoirs remains weak, which is a necessary condition for working within the framework of Born-Markov approximations.

\subsection{The fourth stroke: adiabatic expansion ($D\rightarrow E$ or $D\rightarrow E^{\prime}$) }
The final stroke of the Otto cycle is again an adiabatic one, specifically referred to as an adiabatic expansion stroke. In this path, the frequency of the driving magnetic field is gradually varied from $\omega_{2}$ back to $\omega_{1}$, effectively reversing the process from the second stroke. The final state of the system after completing all four strokes is denoted as 
\begin{equation}
\tilde{\rho}=U^{\dagger}\rho_{D}U.
\end{equation}
During this process, the state evolves from $\rho_{D}$ to $\tilde{\rho}$, and the magnetization at the final point of the cycle remains constant at $S_3$, for the same reason as in the second stroke. In an ideal quantum Otto engine, the final state $\tilde{\rho}$ equals $\rho_A$, and the quantum Otto cycle becomes complete. However, in this paper, during the two isochoric paths of the Otto cycle, we intentionally truncate the processes before reaching equilibrium at a finite time $\tilde{t}$. Consequently, the final state of the system will not be the same as $\rho_A$, and the final magnetization of the system differs from the initial one, i.e., $S_3\ne S_1$, indicating an incomplete Otto cycle.
In this scenario, there are two possibilities: the magnetization at the final point may be greater than the initial magnetization of the system, i.e., $S_3> S_1$, where the final point is denoted as $E$ as shown in Fig.~\ref{fig:fig1}-(a), or the magnetization at the final point will be less than that of the initial, such as $S_3< S_1$, with the final point labeled as $E^{\prime}$ as shown in Fig.~\ref{fig:fig1}-(b). Importantly, the heat transfer during this process is zero, and the amount of work done on the system is given by 
\begin{equation}
W_{2}=\text{Tr}[H_{\mathcal{S}_2}\rho_{D}-H_{\mathcal{S}_1}\tilde{\rho}].
\end{equation}
Therefore, during these four consecutive strokes, a finite-time quantum Otto engine extracts $Q_1$ amount of heat from a hot reservoir of temperature $T_h$, performs $W=W_1+W_2$ amount of work, and releases $Q_2$ amount of heat to a cold reservoir of temperature $T_c$.
\section{Efficiency of  finite-time quantum Otto engine}
\label{sec:3}
The efficiency of a quantum Otto engine is defined as the ratio of the total work done to the amount of heat extracted from the hot reservoir by the system, expressed as $\eta=W/Q_1$. Therefore, for a finite-time quantum Otto engine, the efficiency is given by 
\begin{equation}
\eta
=\frac{S_{2}-S_{3}}{S_{2}-S_{1}}\Big(1-\frac{\omega_{2}}{\omega_{1}}\Big).
\end{equation}
The efficiency of an ideal quantum Otto engine, which operates in the steady-state regime of the two isochoric strokes, can be directly derived from $\eta$ by setting $S_{1}=S_{3}$, resulting in
 \begin{equation}
 \label{es}
     \eta_{s}=1-\frac{\omega_{2}}{\omega_{1}}.
 \end{equation}
 We will denote this $\eta_s$ as the ``ideal frequency" in the rest of the paper. As mentioned earlier, a finite-time quantum Otto engine can result in an incomplete Otto cycle, leading to two possible scenarios. The first scenario occurs when we start with the state $\rho_A$, which is the canonical equilibrium state of the system at temperature $T_c$, as discussed above. In this case, $S_3>S_1$.
 The second situation arises when we start with a state $\tilde{\rho}_{A}=\frac{e^{-\beta^{\prime}H_{\mathcal{S}_{2}}}}{Z_2^{\prime}}$, where $\beta^{\prime}<\beta_c$ and $Z_2^{\prime}=-\text{Tr}(\beta^{\prime}H_{\mathcal{S}_2})$. In this scenario, we obtain $S_3<S_1$.
 In the next section, we will delve into the efficiencies achieved in these two situations. 
 Additionally, we will also explore the impact of a transverse field on this finite-time quantum Otto engine.
\subsection{Efficiency analysis of a finite-time quantum Otto engine: starting from cold bath equilibrium}
\begin{figure*}
     \centering
     \includegraphics[width=8.5cm,height=5.5cm]{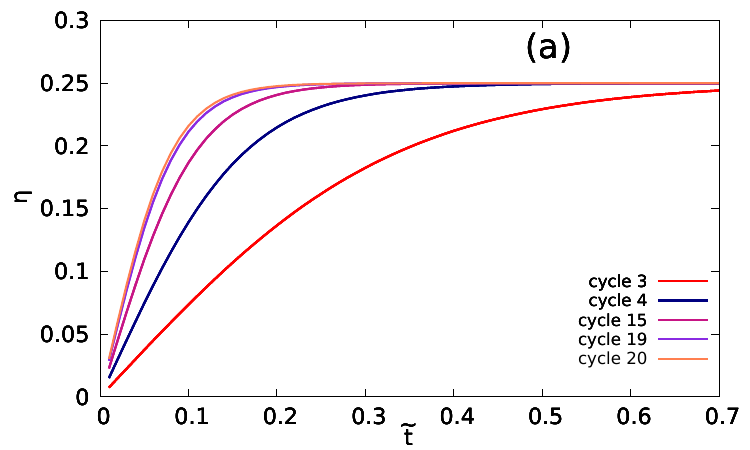}
     \includegraphics[width=8.5cm,height=5.5cm]{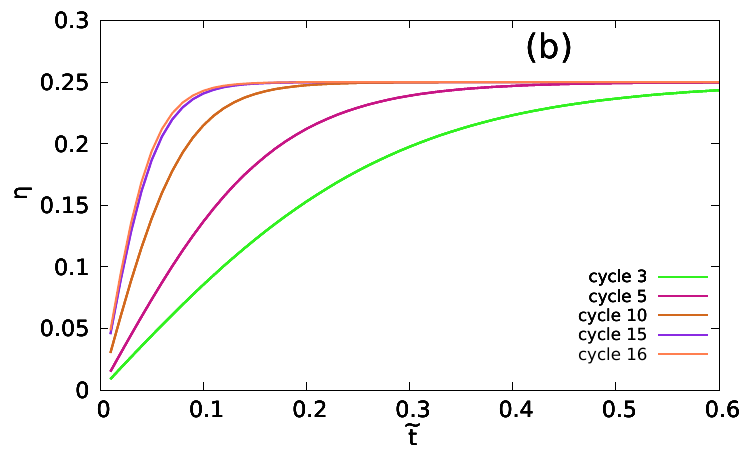}\\
     \vspace{0.5cm}
     \includegraphics[width=8.5cm,height=5.5cm]{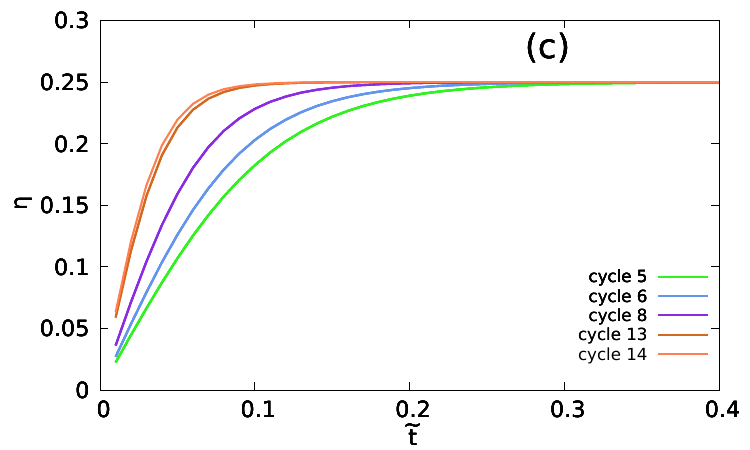}
     \caption{Efficiency in consecutive cycles of a finite-time quantum Otto engine vs. the time duration of isochoric strokes when starting from equilibrium with the cold bath. Here we depict $\eta$ with $\tilde{t}$ by varying magnetic field strengths and temperatures of the cold and hot reservoirs, for (a) 
$\omega_{1}=8.0\,\tilde{\omega}$, $\omega_{2}=6.0\,\tilde{\omega}$, $T_{h}=1.0\frac{\hbar\tilde{\omega}}{k_B}$, and $T_{c}=0.25\frac{\hbar\tilde{\omega}}{k_B}$, (b) $\omega_{1}=16.0\,\tilde{\omega}$, $\omega_{2}=12.0\,\tilde{\omega}$, $T_{h}=2.0\frac{\hbar\tilde{\omega}}{k_B}$, $T_{c}=0.5\frac{\hbar\tilde{\omega}}{k_B}$, and (c) $\omega_{1}=24.0\,\tilde{\omega}$, $\omega_{2}=18.0\,\tilde{\omega}$, $T_{h}=3.0\frac{\hbar\tilde{\omega}}{k_B}$, and $T_{c}=0.75\frac{\hbar\tilde{\omega}}{k_B}$. The different-colored curves represent the efficiency of the finite-time quantum Otto engine for different cycles. The orange curve in each plot represents the limit cycle, which occurs at the $20^{\text{th}}$ cycle for panel (a), the $16^{\text{th}}$ cycle for panel (b), and the $14^{\text{th}}$ cycle for panel (c).  All the quantities plotted here are dimensionless.}
     \label{lc1}
 \end{figure*}



 We now explore the scenario where we initialize the quantum Otto engine with the spin-$1/2$ working substance in the state $\rho_A$ and perform the four strokes as previously discussed for the first cycle. For an ideal quantum Otto engine, after the first cycle we end up with the state, same as the initial state, and can continue repeating the cycle of the engine, which are all identical. Therefore, the efficiency remains the same for all cycles in an ideal engine. However, in the case of a finite-time quantum Otto engine, although we start with the same initial state as an ideal engine, we end up with a state that differs from the initial one after each cycle. The second cycle commences with the state obtained at the end of the first cycle, resulting in different initial state. This pattern continues, with each new cycle starting with the final state of the previous cycle, diverging from the initial state. As a result, the efficiencies of successive cycles differ.
 This process repeats, and after a certain number of cycles, the quantum Otto engine reaches a point where further cycle iterations do not alter its efficiency for a fixed time. This cycle is known as the limit cycle~\cite{PhysRevE.70.046110,feldmann2004characteristics,PhysRevE.94.012119,shirai2021non}. The speed at which we approach the limit cycle depends on the parameters of the driving Hamiltonian, particularly the strength of the magnetic field.
 
 In Fig.~\ref{lc1}, we  illustrate the time dynamics of the efficiencies for consecutive cycles of a finite-time quantum Otto engine
 by varying the strengths of the magnetic fields and the temperatures of the hot and cold reservoirs. As mentioned before, the initial cycle starts with the system in thermal equilibrium at temperature $T_c$.
We observe that increasing the magnetic field strength by the same factor while keeping their ratio $\frac{\omega_1}{\omega_2}$ constant leads to faster convergence towards the limit cycle. This trend is evident when comparing the three panels in Fig.~\ref{lc1}. 
Note that, altering the temperatures while maintaining a constant ratio of $\frac{T_h}{T_c}$ does not significantly impact the efficiency dynamics. Instead, it only helps to increase the magnitude of the heat and work. In the panels of Fig.~\ref{lc1}, we adjust the temperatures in order to maintain a good numerical precision in obtaining the quantities $Q_1$ and $W$. However, if we were to keep the temperatures constant in all three panels, the qualitative nature of the efficiency dynamics would remain the same.
In the specific cases presented in Fig.~\ref{lc1}, it takes $20$ cycles to reach the limit cycle when the strengths of the applied magnetic fields are $\omega_1=8.0\,\Tilde{\omega}$ and $\omega_2=6.0\,\Tilde{\omega}$ (see panel (a)). Conversely, for higher magnetic field strengths, such as $\omega_1=16.0\,\Tilde{\omega}$, $\omega_2=12.0\,\Tilde{\omega}$ in panel (b) and $\omega_1=24.0\,\Tilde{\omega}$, $\omega_2=18.0\,\Tilde{\omega}$ in panel (c), the efficiency of the quantum Otto engine converges after $16$ and $14$ repetitions, respectively. In all the cases presented in Fig.~\ref{lc1}, the final magnetization of the system, $S_3$, after each cycle exceeds the initial magnetization, $S_1$, and for the parameters chosen in the three panels, the efficiency of an ideal quantum Otto engine turns out to be $\eta_s=0.25$, as obtained from Eq.~(\ref{es}). Let us denote the time duration of the isochoric strokes to attain the ideal efficiency $\eta_s$ as $\tilde{t}_s$. Therefore, we can see that for each case depicted in Fig.~\ref{lc1}, the efficiency $\eta$ at time $\tilde{t}<\tilde{t}_s$ is less than $\eta_s$. Hence, in this scenario, a finite-time quantum Otto engine can not achieve a higher efficiency than an ideal one. 
     
  
  \begin{figure}
     \centering
     \includegraphics[width=\linewidth]{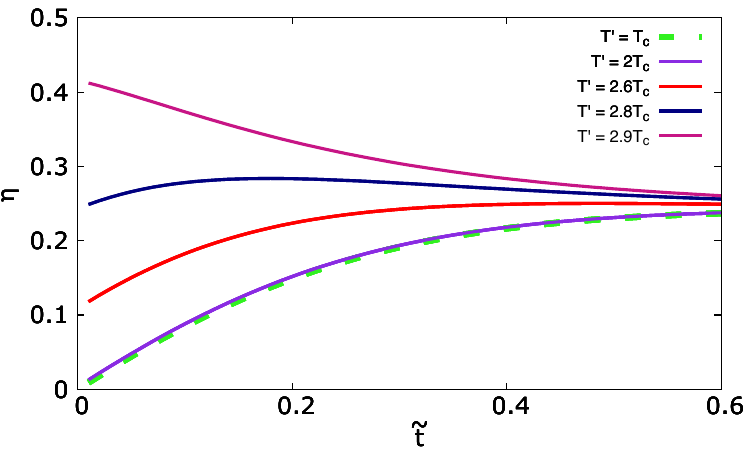}
     \caption{Efficiency of the second cycle of a finite-time quantum Otto engine ($\eta$) vs. the time duration of isochoric strokes $\tilde{t}$ with varying temperature of the initial state of the working substance. 
     The curves are color-coded to represent different initial temperatures of the working substance, with the green line corresponding to the case where the initial temperature is equal to the temperature of the cold bath. All other considerations are same as in Fig.~\ref{lc1}-(c). All the quantities plotted along the horizontal and vertical axes are dimensionless.}
     \label{fig:Ti}
 \end{figure}
\subsection{Efficiency analysis of finite-time quantum Otto engine by tuning the initial state temperature}
\label{IIIB}
In this section, we investigate the efficiency of finite-time quantum Otto engines that start with an initial state in thermal equilibrium with a reservoir at a temperature $T^{\prime}$, where $T_h > T^{\prime} > T_c$, corresponding to the Hamiltonian $H_{\mathcal{S}_2}$. For an ideal quantum Otto engine, if we start with such a state and use the same strokes as discussed in the previous sections, the first cycle will end with the system in thermal equilibrium at temperature $T_c$, corresponding to the Hamiltonian $H_{\mathcal{S}_2}$. This state is different from the initial state of the first cycle. So, the first cycle will be incomplete, and the second cycle will start with the thermal equilibrium state at temperature $T_c$, corresponding to the Hamiltonian $H_{\mathcal{S}_2}$. From the second cycle onward, each cycle of an ideal Otto engine becomes closed, providing the same efficiency as conventional Otto engines. Hence, to ensure cycle closure for an ideal quantum Otto engine counterpart of a finite-time quantum Otto engine, we analyze the engine starting from the second cycle. 
The dependence of efficiency of a finite-time quantum Otto engine on $\tilde{t}$ in this particular case is depicted in Fig.~\ref{fig:Ti} for various initial temperatures of the working substance. We find that, with the increase of $T^{\prime}$ from $T_c$, the efficiency of the second cycle of a finite-time quantum Otto engine also increases, and for certain higher values of $T^{\prime}$, the engine achieves a better efficiency than that of an ideal quantum Otto engine for $\tilde{t}<\tilde{t}_s$. Note that, this increase in efficiency for a short duration of isochoric strokes persists only for a few cycles, and as the number of cycles increases, the efficiency values for a fixed $\tilde{t}$, for $\tilde{t}<\tilde{t}_s$, gradually decrease and ultimately converges to that of the limit cycle.
As time progresses, the efficiency curves for different $T^{\prime}$ values converge to $\eta_s$, the efficiency of an ideal quantum Otto engine as in the previous case shown in Fig.~\ref{lc1}. 
Thus, by adjusting the initial temperature of the working substance, 
the engine's performance can be improved significantly in the far-from-equilibrium region of the two isochoric strokes. Notably, for these finite-time quantum Otto engines, which initiate from a canonical equilibrium state at temperature $T^{\prime}$, we get $S_3<S_1$ exclusively for those specific values of $T^\prime$ where we obtain an advantage over the ideal quantum Otto engine.
An essential observation to make here is that for this case, even after the first cycle of an ideal quantum Otto engine, we still achieve greater efficiency than when starting from an equilibrium state at the cold bath temperature $T_c$. This increase in efficiency for an ideal quantum Otto engine during the first cycle is not unexpected. This phenomenon is a consequence of starting with a state of high energy, characterized by a temperature higher than the cold bath temperature $T_c$. However, what is truly remarkable is that this efficiency gain persists throughout the transient region, even during the second and subsequent cycles. Also, it is important to highlight that one cannot indefinitely raise the temperature $T^{\prime}$, as if it approaches close to the temperature of the hot bath, the heat input $Q_1$ becomes zero, resulting in a situation that lacks a well-defined physical meaning.  
\subsection{Alteration of the driving Hamiltonian with a transverse field}
\begin{figure}
     \centering
     \includegraphics[width=\linewidth]{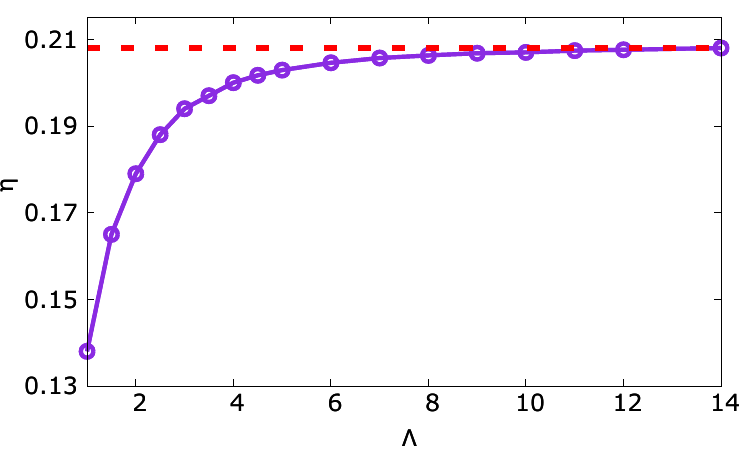}
     \caption{Efficiency of a finite-time quantum Otto engine ($\eta$) in presence of a transverse magnetic field, with the varying magnetic field strength ($\Lambda$). The violet curve with circular points represents the efficiency of a finite-time quantum Otto engine, where both isochoric strokes are truncated at $\tilde{t}=1.0$. The red dashed line represents the constant efficiency value at $\tilde{t}=1.0$ with no transverse field. All other parameters remain the same as in Fig.~\ref{lc1}-(a). All quantities plotted are dimensionless.}
     \label{fig:stray}
 \end{figure}

Up to this point, we have been discussing a quantum Otto engine with finite-time isochoric strokes and perfectly adiabatic strokes, where the adiabatic strokes are governed by the unitary evolution of the system corresponding to the driving Hamiltonian $H_{\mathcal{S}}(t)$ given in Eq.~(\ref{driving1}).  We have found that, in certain cases depending on the temperature of the initial canonical equilibrium state of the working substance, we can enhance the efficiency of a finite-time quantum Otto engine beyond that of an ideal one. In this section, our goal is to study the effect of a transverse field on the efficiency of a finite-time quantum Otto engine. For this investigation, we start with the state $\rho_A$ defined previously. We introduce a transverse field along the $x$ direction in the driving Hamiltonian $H_{\mathcal{S}}(t)$ and analyze the efficiency of the engine while varying the strength of the applied field. The driving Hamiltonian during the two adiabatic branches hence can be written as
 \begin{equation}
 \Tilde{H}_{\mathcal{S}}(t)=\frac{1}{2}\hbar\omega(t)\sigma_z+\hbar\xi\\(t)\sigma_x,
     \end{equation}
where at time $t=0$, $\tilde{H}_{\mathcal{S}}(0)=\tilde{H}_{\mathcal{S}_1}=\frac{1}{2}\hbar\omega_{1}\sigma_{z}+\xi_{1}\sigma_{x}$ and at time $t=\tau$, $\tilde{H}_{\mathcal{S}}(\tau)=\tilde{H}_{\mathcal{S}_2}=\frac{1}{2}\hbar\omega_{2}\sigma_{z}+\xi_{2}\sigma_{x}$. 
For a perfectly adiabatic stroke, attained by the protocol outlined in Section~\ref{sec:2.B}, in case of a finite-time quantum Otto engine, it is necessary to satisfy the condition $[\tilde{H}_{\mathcal{S}_1},\tilde{H}_{\mathcal{S}_2}]=0$. To meet this requirement, we take 
$\xi(t)=\frac{\omega(t)}{\Lambda}$, where $\Lambda$ is a dimensionless constant. Consequently, we have $\xi_{1}=\frac{\omega_{1}}{\Lambda}$ and $\xi_{2}=\frac{\omega_{2}}{\Lambda}$. Utilizing the driving Hamiltonian $\tilde{H}_{\mathcal{S}}(t)$ and this specific choice of $\xi(t)$, we can derive the efficiency of the ideal quantum Otto engine as
\begin{equation}
\tilde{\eta}_s=1-\frac{\omega_{2}(S_{2}-S_{1})+2\xi_{2}(\tilde{S}_{2}-\tilde{S}_{1})}{\omega_{1}(S_{2}-S_{1})+2\xi_{1}(\tilde{S}_{2}-\tilde{S}_{1})},
\label{effin}
\end{equation}
where,
$\tilde{S}_{2}=\text{Tr}[\frac{1}{2}\sigma_{x}\overline{\rho}_{B}]=\text{Tr}[\frac{1}{2}\sigma_{x}\overline{\rho}_{C}]$ and
$\tilde{S}_{1}=\text{Tr}[\frac{1}{2}\sigma_{x}\rho_{A}]=\text{Tr}[\frac{1}{2}\sigma_{x}\overline{\rho}_{D}]$. Here, $\overline{\rho}_B$, $\overline{\rho}_C$, and $\overline{\rho}_D$ are the states of the system at points $B$, $C$ and $D$, respectively. Now, by substituting the values of $\xi_1$ and $\xi_2$, we find that $\tilde{\eta}_s=\eta_s$. Therefore, even in the presence of the transverse field, the efficiency of an ideal quantum Otto engine can be equal to that of the case without a transverse field, as given in Eq.~(\ref{es}), if $\xi_{1}=\frac{\omega_{1}}{\Lambda}$ and $\xi_{2}=\frac{\omega_{2}}{\Lambda}$.
We now investigate how the efficiency of the finite-time quantum Otto engine depends on the strength of the transverse field. Fig.~\ref{fig:stray} illustrates the variation in efficiency with changes in $\Lambda$ for a fixed transient time $\tilde{t}=1.0$ of the two isochoric strokes. Here, we have chosen $\tilde{t}$ to be one-fourth of $\tilde{t}_s$. It is evident from the figure that as the strength of the transverse field increases or the value of $\Lambda$ decreases, the efficiency of a finite-time quantum Otto engine decreases. Moreover, a finite-time quantum Otto engine with no transverse field provides better efficiency than the one with a transverse field. As the field strength decreases, the efficiency of the finite-time quantum Otto engine increases and ultimately saturates to the value of efficiency with no transverse field. Therefore, applying a transverse field is not beneficial for enhancing the efficiency of a finite-time quantum Otto engine.

\section{Incorporation of  auxiliary qubit with working substance}
\label{sec:4}
\begin{figure}
     \centering
     \includegraphics[width=\linewidth]{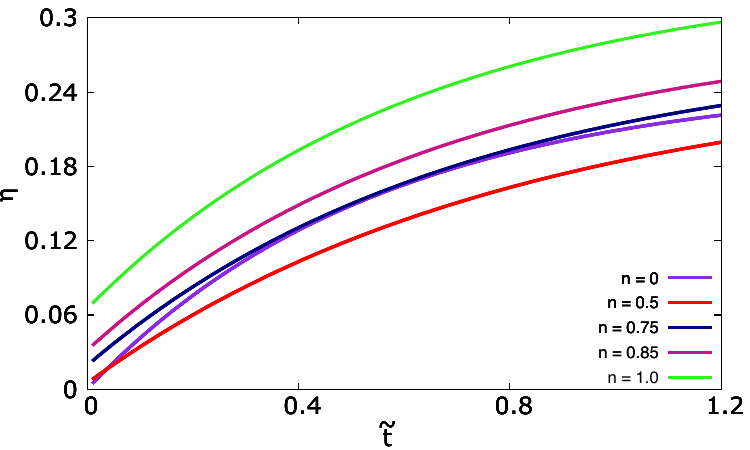}
     \caption{Efficiency of a finite-time quantum Otto engine ($\eta$) vs. the time duration of isochoric strokes ($\tilde{t}$) by varying the interaction strength between the system and the auxiliary qubit. 
     Here the bath temperatures are taken to be, $T_{h}=0.5\frac{\hbar\omega}{k_B}$, $T_{c}=0.25\frac{\hbar\omega}{k_B}$. The values of $\omega_1$ and $\omega_2$ are same as in Fig.~\ref{lc1}-(a). The different colors of the curves denote different values of $n$. 
     All the quantities plotted here are dimensionless.}
   \label{int}
 \end{figure}
In the previous section, we examined the efficiency of a finite-time quantum Otto engine with a spin-$1/2$ particle as the working substance in various scenarios. In this section, we investigate the influence of introducing an auxiliary qubit alongside the single-qubit working substance. For this scenario, we execute the four strokes of the quantum Otto engine by taking into account the additional auxiliary qubit with the working substance.
However, in the two isochoric strokes, the reservoirs are connected only to the system qubit (the working substance), not to the auxiliary qubit, with the interaction Hamiltonian being
 \begin{equation}
 H^\prime_{\mathcal{SB}_i}=\sigma_{x}\otimes I_{\mathcal{A}} \sum_{k}\hbar g^i_{k}(a_{k}^{i\dagger}+a_{k}^i).
 \end{equation}
Here, $i=h$ or $c$ for the hot and cold reservoirs, respectively, and $I_{\mathcal{A}}$ is the identity operator on the Hilbert space of the auxiliary qubit. So, the total Hamiltonian of the composite setup in the isochoric strokes is given by
 \begin{equation}
 H^{\prime}=H_{\mathcal{S}_j}\otimes I_{\mathcal{A}}+I_{\mathcal{S}} \otimes H_{\mathcal{A}_j}+H_{I_j}+H_{\mathcal{B}_i}+H^{\prime}_{\mathcal{SB}_i}.
 \end{equation}
 Here, $I_{\mathcal{S}}$ represents the identity operator on the Hilbert space of the system qubit. $H_{\mathcal{A}_j}$ is the free Hamiltonian of the auxiliary qubit given by
 $H_{\mathcal{A}_j}=\frac{1}{2}\hbar\omega_j \sigma_z^{\mathcal{A}}$. The superscript ``$\mathcal{A}$" in $\sigma_z^{\mathcal{A}}$ indicates that the $\sigma_z$ operator is defined in the Hilbert space of the auxiliary qubit.
 The interaction between the system and the auxiliary qubit is taken as 
 \begin{equation}
 H_{I_j}=\varepsilon_j (\sigma_{x}\otimes \sigma_{x}^\mathcal{A}),
 \end{equation}
 with $\varepsilon_j$ being the coupling strength of the system-auxiliary interaction. In the above expressions, $j=1$ when $i=h$ and $j=2$ when $i=c$.
 For the adiabatic strokes, the driving Hamiltonian is taken as
 \begin{equation}
 \label{Hin1}
 H^{\prime}_{\mathcal{SA}}(t)=H_{\mathcal{S}}(t)\otimes I_{\mathcal{A}}+I_{\mathcal{S}} \otimes H_{\mathcal{A}}(t)
 +\varepsilon(t)(\sigma_{x}\otimes\sigma_{x}^\mathcal{A}),
 \end{equation}
 where $H_{\mathcal{A}}(t)=\frac{1}{2}\hbar\omega(t)\sigma_z^{\mathcal{A}}$ and $\varepsilon(t)=n\hbar\omega(t)$, with $n$ being a real number. At $t=0$, the Hamiltonian of the composite system-auxiliary setup becomes $H^{\prime}_{\mathcal{SA}_1}=H^{\prime}_{\mathcal{SA}}(0)$, with $\omega(0)=\omega_1$, and at $t=\tau$, with $\omega(\tau)=\omega_2$, the Hamiltonian becomes $H^{\prime}_{\mathcal{SA}_2}=H^{\prime}_{\mathcal{SA}}(\tau)$. Now, starting with the composite system-auxiliary state $\rho^{\prime}_{A}=\frac{e^{-\beta_{c}H^{\prime}_{\mathcal{SA}_2}}}{\text{Tr}(e^{-\beta_{c}H^{\prime}_{\mathcal{SA}_2}})}$, at point $A$, we perform the consecutive four strokes described previously and the heat taken and released by the system qubit, after tracing out the auxiliary, is given by
\begin{eqnarray}
    &&Q^{\prime}_{1}=\text{Tr}(H_{\mathcal{S}_1}(\text{Tr}_\mathcal{A}(\rho^{\prime}_{B}-\rho^{\prime}_{A})))\nonumber\\
   \text{and} \quad && 
   Q^{\prime}_{2}=\text{Tr}(H_{\mathcal{S}_2}(\text{Tr}_\mathcal{A}(\rho^{\prime}_{D}-\rho^{\prime}_{C}))),
\end{eqnarray}
respectively, where $\rho^{\prime}_B$, $\rho^{\prime}_C$, and $\rho^{\prime}_D$ are the composite system-auxiliary states at points $B$, $C$ and $D$, respectively. $\text{Tr}_\mathcal{A}$ is the partial trace taken on the auxiliary qubit. In this case also, $[H^{\prime}_{\mathcal{SA}_1},H^{\prime}_{\mathcal{SA}_2}]=0$, hence we perform the adiabatic stroke in the same way as in the previous scenarios. 
Thus the work done by the system qubit in the adiabatic compression and adiabatic expansion strokes can be defined as
\begin{eqnarray}
&&W^{\prime}_{1}=\text{Tr}[H_{\mathcal{S}_1}\text{Tr}_\mathcal{A}(\rho^{\prime}_{B})-H_{\mathcal{S}_2}\text{Tr}_\mathcal{A}(\rho^{\prime}_{C})]\nonumber\\
\text{and}\quad &&W^{\prime}_{2}=\text{Tr}[H_{\mathcal{S}_2}\text{Tr}_\mathcal{A}(\rho^{\prime}_{D})-H_{\mathcal{S}_1}\text{Tr}_\mathcal{A}(\rho^{\prime})],
\end{eqnarray}
respectively, with $\rho^{\prime}$ being the final state of the composite system-auxiliary qubit after the four strokes.
 \begin{figure*}
     \centering
     \includegraphics[width=5.9cm, height=5cm]{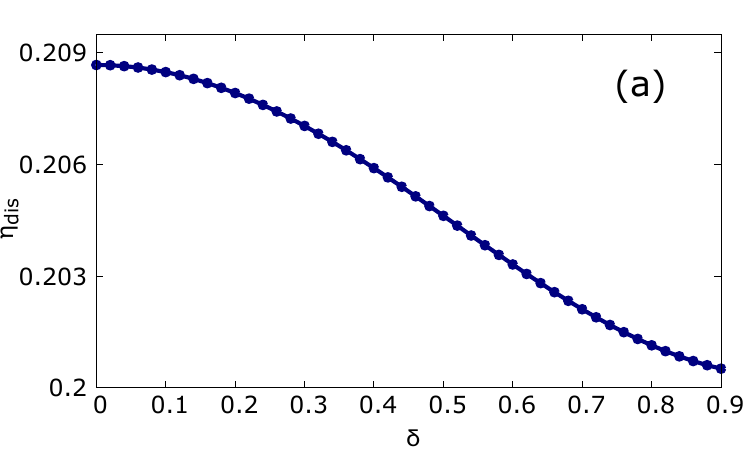}
     \includegraphics[width=5.9cm, height=5cm]{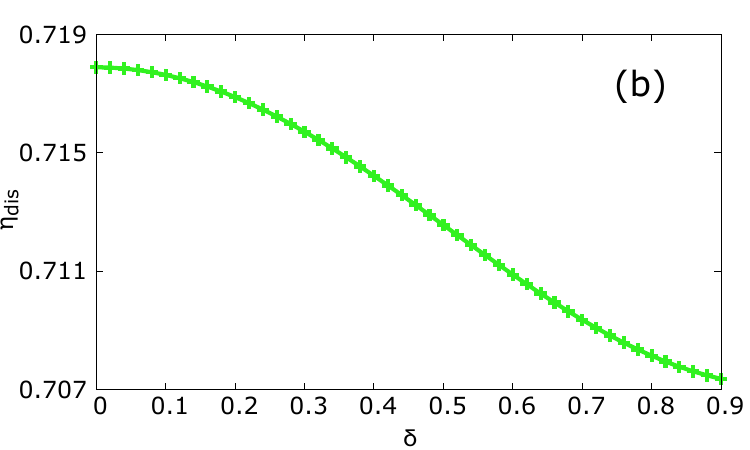}
      \includegraphics[width=5.9cm, height=5cm]{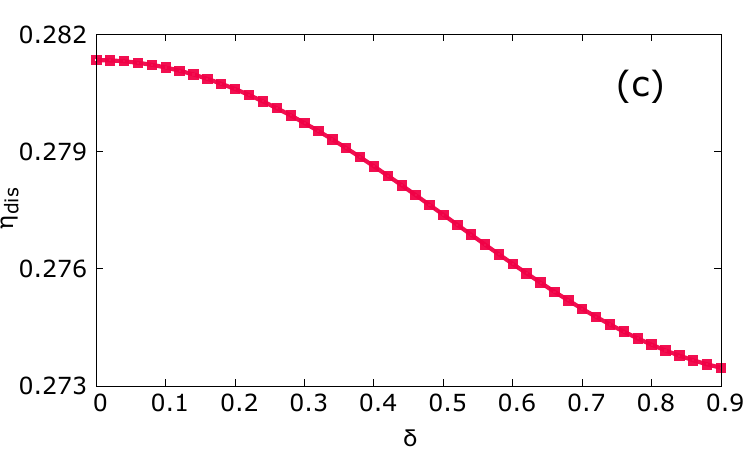}
     \caption{Effects of discrete glassy disorder present in the system-bath coupling parameter on the efficiency of a finite-time quantum Otto engine. Here we plot the efficiency of the finite-time quantum Otto engine, $\eta_{\text{dis}}$, with the strength of disorder, $\delta$, (a) when we start from a canonical equilibrium state of the working substance at the cold bath temperature $T_c$, i.e., $\rho_A$, (b) when we start from a canonical equilibrium state of the working substance at a temperature $T^{\prime}=2.15\frac{\hbar\omega}{k_B}$, where $T_h>T^{\prime}>T_c$, i.e., $\tilde{\rho}_A$, and (c) in presence of an auxiliary qubit with the working substance, i.e., the initial state is $\rho^{\prime}_A$. Here we set $n=1.0$. For panels (a) and (b) the strengths of the magnetic field and the temperature of the hot and cold reservoirs are same as in Fig.~ 
     \ref{lc1}-(c), while for panel (c) these considerations are same as in 
     Fig.~\ref{int}. For all the three panels, the probability of the occurrence of the ordered scenario is kept as $p=0.9$ and the transient time for the two isochoric strokes is set to be 
     $\Tilde{t} =1.0$. All the quantities plotted here are dimensionless.
     \label{disorder}}.
 \end{figure*}
 

We now vary the value of $n$ and, thereby, the interaction strength between the system-auxiliary qubits, to observe how the interaction strength affects the efficiency of a finite-time quantum Otto engine. Interestingly, as seen in Fig.~\ref{int}, for a small value of $n$ (e.g., $n=0.5$), the efficiency of the finite-time quantum Otto engine is lower than that observed in a scenario where there is no interaction between the system and the auxiliary qubit, effectively representing the absence of the auxiliary qubit, as denoted by the $n=0$ curve. However, with the increase of $n$ from $0.75$ to $1.0$, the efficiency improves progressively and eventually surpasses the efficiency values obtained with 
$n=0$ line. This indicates that the inclusion of an additional auxiliary qubit within the working substance can enhance the efficiency of a finite-time quantum Otto engine compared to the scenario without an auxiliary qubit. Note that, this efficiency enhancement through the incorporation of an auxiliary qubit is also applicable to ideal quantum Otto engines. Specifically, for $n=0.75$ to $1.0$, we can achieve higher efficiency compared to $\eta_s$ for an ideal quantum Otto engine. It is also important to mention that we cannot increase the value of $n$ indefinitely, as exceeding $n=1.0$ would violate the weak-coupling condition required for the validation of Born-Markov approximations in the two isochoric strokes.


\section{Effects of glassy disorder on efficiency of finite-time quantum Otto engine}
\label{sec:5}
 Disorder refers to the presence of uncertainties within a system. In the real world, no system can be considered perfect; there will always be inherent uncertainties associated with various system parameters. Interestingly, these uncertainties can occasionally yield advantageous outcomes, leading to improved performance when compared to ordered scenarios. In this section, we will conduct an analysis of the impact of disorder, present
 in the coupling parameter of the system and its environment, which modifies the decay constant denoted as $\gamma_i(\mathcal{E})$, defined in Eq.~(\ref{gamma}), and consequently affects the overall efficiency.

 In this paper, we consider the disordered parameters to be glassy. Glassy disorder in a system is characterized by the presence of a disordered system parameter, with an equilibration time significantly longer than the relevant time scales for our investigation of the system. This means that during our observation period, the disordered parameters remain effectively unchanged for a specific realization of disorder. This concept of disorder is analogous to that observed in spin glass systems ~\cite{d1,d2}. Glassy disorder is also referred to as ``quenched disorder" in the literature ~\cite{d3,d1,d2,d4,d5}. Studies exploring spin chains with glassy disorder have been conducted and documented in ~\cite{d6,d7,d8,d9,d10,d11,d12,d13}.
 
We now investigate how  disorder in the coupling strength of the system and the reservoirs can manipulate  engine's efficiency. We introduce a disorder parameter in the system-bath interaction Hamiltonian, given in Eq.~(\ref{int}), as
  \begin{equation}
H_{\mathcal{SB}_{i_{\text{dis}}}}=\sigma_{x}\sum_{k}\hbar g^i_{k}(1+d_i)(a_{k}^{i\dagger} + a_{k}^i).
  \end{equation}
  Here $d_i$ is the disorder parameter which we choose randomly from a discrete probability distribution function given as
   \begin{eqnarray}
    P(d_i) &=& \frac{(1-p)}{2}\quad \text{when} \quad d_i=\pm\delta, \nonumber \\
                &=& p  \quad \quad \phantom{tum} \text{when} \quad d_i=0, \nonumber \\ 
               &=& 0 \; \; \; \; \; \;\;\;\;\;\;\;\;\;  \text{otherwise}, 
    \label{eq:discrete}
    \end{eqnarray}
  with $\delta>0$ and $0\le p<1$. As a consequence of the presence of this type of disorder, the transition rates, given in Eq.~(\ref{gamma}), are altered as
\begin{align}
\label{gamma_dis}
\gamma_{i_{\text{dis}}}(\mathcal{E})=
    \begin{cases}
        \mathcal {J}_i(\mathcal{E})(1+d_i)^2(1+\overline{n}_i(\mathcal{E})), & \mathcal{E} > 0,\\
        \mathcal {J}_i(|\mathcal{E}|)(1+d_i)^2(\overline{n}_i(|\mathcal{E}|), & \mathcal{E} < 0,
    \end{cases}
\end{align}
for one realization of disorder. Here $i=h$ for isochoric heating and $i=c$ for isochoric cooling stroke. The disorder averaged efficiency of a finite-time quantum Otto engine is given by
 \begin{equation}
     \eta_{\text{dis}}=\sum_{j}p_{j}\eta_{j}.
 \end{equation}
 Here $j$ runs from $1$ to $3$ corresponding to the $d_i$ values being $\delta$, $0$, and $-\delta$ respectively, with the corresponding probabilities $p_j$.
  We consider situation when 
  the probability of occurrence of the disordered situation is very low, i.e., the ordered case is more probable with probability $p=0.9$. 
  In Fig.~\ref{disorder} we depict the effects of this discrete type disorder on the efficiency of a finite-time quantum Otto engine. In panel (a), we present the situation where we have a spin-$1/2$ particle as the working substance of the engine and we start the engine by preparing the initial state of the working substance at the equilibrium with the cold bath, i.e., the state $\rho_A$. In this case, we keep the transient time of the two isochoric strokes fixed at $\tilde{t}=1.0$. We observe that in the ordered scenario with $\delta=0$, the efficiency of the finite-time quantum Otto engine is at its maximum, and as the disorder strength $\delta$ increases, the efficiency gradually decreases. Importantly, this reduction in efficiency with increasing disorder strength is not very rapid.

  From the previous discussions of this paper, we have seen that by tuning the temperature of the initial state in the temperature range between the hot and the cold baths we can achieve a better efficiency of a finite-time quantum Otto engine compared to that of an ideal one. We now investigate whether this advantageous scenario persists even in the presence of glassy disorder within the system-bath coupling parameter. For the case depicted in Fig.~\ref{disorder}-(b), we initiate the engine with the working substance prepared in a canonical equilibrium state at a temperature $T^{\prime}=2.15\frac{\hbar\omega}{k_B}$, which is higher than the cold bath temperature $T_c=0.75\frac{\hbar\omega}{k_B}$ and lower than the hot bath temperature $T_h=3.0 \frac{\hbar\omega}{k_B}$. We observe that, here also the influence of discrete-type glassy disorder remains qualitatively same as in the previous case depicted in panel (a). However, it is crucial to note that with a substantial amount of disorder strength $\delta$, the efficiency of the finite-time quantum Otto engine exceeds that of an ordered ideal quantum Otto engine, for which the efficiency, $\eta_s=0.25$, with the system and bath parameters considered in this depiction. Hence, tuning the temperature of the initial state of the working substance remains beneficial even in the presence of the discrete glassy disorder examined in this paper.

  Next we incorporate an auxiliary qubit with the working substance as in Sec.~\ref{sec:4}, and begin with the composite state of the system-auxiliary, $\rho^{\prime}_{A}$. In this case as well, disorder has a detrimental effect on the efficiency of a finite-time quantum Otto engine, but the efficiency with a sufficiently large value of the disorder strength is higher than the efficiency of an ordered quantum Otto engine operating without the auxiliary qubit. See Fig.~\ref{disorder}-(c). Note that, in all the panels of Fig.~\ref{disorder}, we have demonstrated only the first cycle of a finite-time quantum Otto engine. 

\begin{figure}
     \centering
     \includegraphics[width=8.5cm, height=5.5cm]{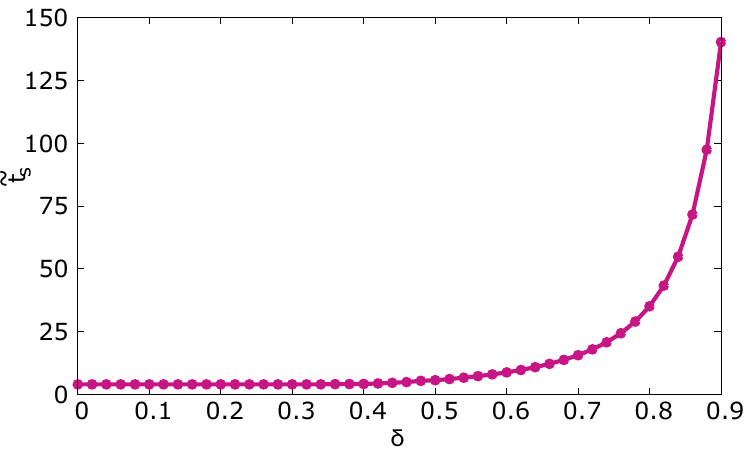}
     \caption{Variation of ideal efficiency reaching time with the increase of disorder strength. Here we depict $\tilde{t}_{s}$ vs. the disorder strength $\delta$ for $p=0.9$. The initial state of the system is taken as $\rho_A$. The other considerations are same as in Fig.~\ref{lc1}-(c). The quantities plotted along the horizontal and the vertical axes are dimensionless.}
     \label{fig:steady}
 \end{figure}
  
Till now, we have investigated how disorders affect the efficiency of finite-time quantum Otto engines. However, we are now shifting our attention to explore the impact of such disorders on the efficiency of an ideal quantum Otto engine. Based on the previous investigations of this paper using a spin-$1/2$ particle as the working substance and starting from an equilibrium setup with the cold bath, we have observed that, the efficiency of an ideal quantum Otto engine, operating in the steady-state regime of the two isochoric strokes, is higher than the efficiency of a finite-time quantum Otto engine. As impurities are ubiquitous in nature, it is essential to analyze the effects of disorder on ideal quantum Otto engines as well. In this study, as we consider disorder in the system-bath coupling parameter and allow the system to equilibrate with the hot and cold bath temperatures during the respective isochoric strokes in the case of an ideal quantum Otto engine, we find that the engine still reaches its ideal efficiency, unaffected by this type of disorder. However, the presence of disorder does impact the time required for the isochoric strokes, $\tilde{t}_s$, to converge to the efficiency of an ideal quantum Otto engine, $\eta_s$.
In Fig.~\ref{fig:steady}, we illustrate the variation of $\tilde{t}_s$ with increasing disorder strength $\delta$, when starting with the canonical equilibrium state of the system corresponding to the cold bath temperature. No auxiliary qubit is present in this scenario. Here the disorder strength is incremented from $0$ to $0.9$, with the probability of encountering a disordered case being consistently kept at a very low value by setting the probability $p$ to a high value ($p=0.9$). As evident from the figure, for disorder strengths up to $\delta\approx 0.3$, the system exhibits robustness against disorder in terms of the required time duration of the isochoric strokes to attain the ideal efficiency, with $\tilde{t}_{s}\approx 4.0$. Whereas, as the strength of disorder increases, the system progressively slows down, requiring significantly longer times to reach the ideal efficiency $\eta_s$. Throughout this entire discussion involving disorder, we have exclusively considered a discrete-type disorder. Nonetheless, if we randomly choose the disorder parameter $d_i$ from a continuous probability distribution, such as a Normal distribution, the outcomes discussed in this section will exhibit similar qualitative characteristics.
 \section{Conclusion}
 \label{sec:6}
In this paper, we have analyzed the operational characteristics of finite-time quantum Otto engines. This engine deviate from the conventional ideal quantum Otto engines in that they curtail the two isochoric strokes before the working substance reaches thermal equilibriums with the hot and cold baths. Our focus has been on quantum Otto engines that employ a spin-$1/2$ particle as their primary working substance, possibly with an auxiliary.

We observed that despite experiencing incomplete Otto cycles, these single-qubit finite-time quantum Otto engines, under particular circumstances, can outperform their counterparts in terms of efficiency. This efficiency improvement can be achieved by precisely tuning the initial temperature of the working substance within the temperature range defined by the hot and cold baths.
Furthermore, we have also found that the inclusion of an auxiliary qubit, coupled with specific interactions between the single-qubit working substance, can result in efficiency enhancements for both finite-time and ideal quantum Otto engines when compared to scenarios in which the auxiliary qubit is not utilized.

Additionally, we looked at the response to introduction of disorder in the system-bath coupling, particularly during the two isochoric strokes, on the efficiency of finite-time quantum Otto engines. Our findings reveal that as disorder strength increases, the efficiency tends to decrease. However, this reduction remains relatively modest even at high disorder strengths. It is worth noting that the efficiency advantage maintained by finite-time quantum Otto engines over their ideal counterparts by tuning the initial state temperature, is retained even in the presence of significant strengths of disorder. Also, the efficiency improvement observed in finite-time quantum Otto engines when an auxiliary qubit is introduced, as opposed to the setup without the auxiliary one, remains effective even in presence of significant disorder. Our research also confirmed that the presence of such disorder does not alter the ideal efficiency, i.e., the efficiency an ideal quantum Otto engine would achieve. However, it does exert an influence on the duration of isochoric strokes necessary within a quantum Otto engine to reach this ideal efficiency. This duration remains relatively stable until a specific strength of disorder. Beyond this critical threshold, an increase in disorder strength leads to a rapid increase in the required stroke duration.

\acknowledgements 
 We acknowledge computations performed using Armadillo~\cite{Sanderson,Sanderson1}, and QIClib~\cite{QIClib}. This research was supported in part by the `INFOSYS scholarship for senior students'. We also acknowledge partial support from the Department of Science and Technology, Government of India through the QuEST grant (grant number DST/ICPS/QUST/Theme-3/2019/120).
\bibliography{ref} 
\end{document}